\documentclass[mnsc,nonblindrev]{informs3}
\usepackage{algorithmicx}
\usepackage[ruled]{algorithm2e}

\usepackage{amsmath,amsfonts,amssymb,bbm} 
\OneAndAHalfSpacedXI
\newcommand{\me}{\mathrm{e}}
\newcommand{\cK}{\mathcal{K}}

\newcommand{\cR}{\mathcal{R}}

\newcommand{\bp}{\mathbf{p}}

\usepackage{natbib}
 \bibpunct[, ]{(}{)}{,}{a}{}{,}%
 \def\bibfont{\small}%
\TheoremsNumberedThrough     % Preferred (Theorem 1, Lemma 1, Theorem 2)
\ECRepeatTheorems

\EquationsNumberedThrough    % Default: (1), (2), ...
\MANUSCRIPTNO{}

\begin{document}
%%%%%%%%%%%%%%%%

%\setcounter{page}{1}
%\DOI{10.1287/xxxx.0000.0000}%

\RUNAUTHOR{Vijay Kamble}

\RUNTITLE{Revenue Management on an On-Demand Service Platform}

% \TITLE{Bundling Information Goods of Decreasing Value}
\TITLE{Revenue Management on an On-Demand Service Platform}

\ARTICLEAUTHORS{%
\AUTHOR{Vijay Kamble}
\AFF{Department of Information and Decision Sciences\\University of Illinois at Chicago\\ \EMAIL{kamble@uic.edu}} %, \URL{}}
} % end of the block

\ABSTRACT{%
I consider the optimal hourly (or per-unit-time in general) pricing problem faced by a worker (or a service provider) on an on-demand service platform. Service requests arriving while the worker is busy are lost forever. Thus, the optimal hourly prices need to capture the average hourly opportunity costs incurred by accepting jobs. Due to potential asymmetries in these costs, price discrimination across jobs based on duration, characteristics of the arrival process, etc., may be necessary for optimality, even if the customers' hourly willingness to pay is identically distributed. I first establish that such price discrimination is not necessary if the customer arrival process is Poisson: in this case, the optimal policy charges an identical hourly rate for all jobs. This result holds even if the earnings are discounted over time. I then consider the case where the customers belong to different classes that are differentiated in their willingness to pay. I present a simple and practical iterative procedure to compute the optimal prices in this case under standard regularity assumptions on the distributions of customer valuations. I finally show that these insights continue to hold in the presence of competition between multiple quality-differentiated workers, assuming a natural customer choice model in which a customer always chooses the best available worker that she can afford.
% Enter your abstract
}%

%\KEYWORDS{deterministic inventory theory; infinite linear programming duality;
%  existence of optimal policies; semi-Markov decision process; cyclic schedule}

\KEYWORDS{On-Demand Services, Freelancer Pricing, Revenue Management} 
%\HISTORY{This paper was
%first submitted on April 12, 1922 and has been with the authors for
%83 years for 65 revisions.}

\maketitle
%%%%%%%%%%%%%%%%%%%%%%%%%%%%%%%%%%%%%%%%%%%%%%%%%%%%%%%%%%%%%%%%%%%%%%

% Samples of sectioning (and labeling) in MNSC
% NOTE: (1) \section and \subsection do NOT end with a period
%       (2) \subsubsection and lower need end punctuation
%       (3) capitalization is as shown (title style).
%
%\section{Introduction.}\label{intro} %%1.
%\subsection{Duality and the Classical EOQ Problem.}\label{class-EOQ} %% 1.1.
%\subsection{Outline.}\label{outline1} %% 1.2.
%\subsubsection{Cyclic Schedules for the General Deterministic SMDP.}
%  \label{cyclic-schedules} %% 1.2.1
%\section{Problem Description.}\label{problemdescription} %% 2.

% Text of your paper here

\section{Introduction}
The recent years have seen an exponential rise of the so called ``gig economy'',  broadly consisting of online and in-person work relationships that are facilitated ``on-demand'' or ``just-in-time'' by online platforms \citep{de2015rise}. While this paradigm has ushered in new opportunities for flexible employment in the economy, navigating the uncertainties associated with short-term contractual work can be daunting. This paper is motivated by the goal of empowering workers and service providers on these platforms to take effective operational decisions in these complex and uncertain settings. 

In particular, this paper is concerned with the revenue management challenges faced by workers on platforms where they individually make their own pricing decisions. Examples include platforms that match semi-skilled or skilled labor to tasks of heterogeneous nature, like Upwork, Taskrabbit, or Thumbtack, but do not include ride-sharing platforms like Uber and Lyft, where the pricing is centrally regulated by the platform. These latter pricing decisions have received considerable attention from an operational perspective in recent literature (e.g., \citealt{banerjee2015,bimpikis2019spatial,taylor2018demand,cachon2017role,castillo2017surge}). However, despite the presence of several blogs and online forums that informally and anecdotally discuss various aspects of the issue of freelancer pricing,\footnote{A web search for ``freelancer pricing strategies'' returns hundreds of entries on the topic.} it has received little formal attention from the scientific community. To a large extent, similar revenue management challenges are faced by service providers on online rental marketplaces like Airbnb for lodgings, Turo or Getaround for cars, etc.

In this paper, I consider the central problem faced by a worker on an on-demand service platform, which is that of optimally pricing her services to maximize earnings. Because of uncertainty and heterogeneity in the job durations, workers typically use per-unit-time pricing (e.g., hourly), which allows the the payment to scale linearly with the duration of a job.\footnote{Such pricing is standard not only on on-demand labor platforms like Upwork, Taskrabbit, and Thumbtack (hourly), but also on rental marketplaces like Airbnb (per-night) or Turo/Getaround (hourly).} Consequently, the focus is on such per-unit-time pricing strategies (I discuss the relation to per job pricing in Section~\ref{sec:extapp}).

A key feature of the on-demand service economy is that there are no ``per worker'' queues: if a particular worker is unavailable, the customer simply chooses some other worker or leaves the platform. Hence, from the perspective of a worker, accepting a job incurs an opportunity cost of losing all the jobs that could have been accepted while the worker is busy. The optimal price per-unit-time for a job thus needs to internalize the \emph{per-unit-time opportunity cost}, which is the ratio of the expected earnings lost due to accepting the job and the expected duration of the job.

The \emph{first} contribution of this paper is to highlight that, depending on system characteristics and class specific features like distributions of the job durations, particulars of the arrival processes, etc., there could be an asymmetry in the per-unit-time opportunity costs incurred by different classes of jobs. And in these cases, price discrimination across these classes may be necessary to maximize earnings even if there is no difference in the beliefs about the customers' per-unit-time willingness to pay. 
The following example is a bit contrived, but it nevertheless illustrates this point. More examples of practical interest will be presented in Section~\ref{sec:brittle}.

\emph{{\bf Example.} Consider two types of customers: type A customers bring a job of length $1$ hour and type B customers bring a job of length $2$ hours. The willingness to pay per hour for each customer type is uniformly distributed in $[0,1]$. Type $A$ customers arrive at times $t=0,2,4,6,\ldots$ and type B customers arrive at times $t=1,3,5,7,\ldots$. Now, accepting a job of type $A$ does not incur any opportunity cost since it will be finished in time to be able to accept the next job. Thus the optimal hourly price for this job can be independently computed to be $1/2$ per hour. On the other hand, accepting a customer of type $B$ does incur an opportunity cost: the next arriving customer of type $A$ cannot be accepted. This necessitates a reserve value on the hourly price for job type $B$ resulting from the requirement that the total earning from the job is at least as much as the optimal expected earning from the type $A$ job that would be lost. Consequently, the optimal hourly price for job type $B$ is higher than $1/2$.}

The \emph{second} contribution of this paper is to establish that if the customers' arrival process is Poisson, and if their per-unit-time valuations are drawn from an identical distribution, the optimal pricing strategy that maximizes the long-run average earning charges a single rate for all jobs. In other words, in this case, the per-unit-time opportunity costs are identical across jobs irrespective of their durations.  Discounting earnings over time, i.e., giving more importance to the earnings obtained earlier, does not affect this result. Hence, any price discrimination must result from differences in the distributions of the customers' willingness to pay.  

I present this particular finding with some hesitancy, since once the question is appropriately formalized, the result is mathematically trivial. However, it may not be quite as obvious, and could provide useful guidance to workers and service providers as they undertake their pricing decisions. For example, the desire to improve capacity utilization in the face of arrival uncertainty
may tempt a worker to prioritize longer jobs by offering them a lower hourly price as
compared to jobs that are expected to be relatively short, even if the customers' hourly willingness to pay is believed to be identically distributed across jobs.\footnote{For instance, price discounts for longer stays are commonly observed on rental marketplaces like Airbnb.} This temptation may be stronger if the worker is time sensitive and wishes to maximize discounted earnings. The result implies that such intuition is unfounded. 

This brings us to the \emph{third} contribution of the paper.  Under the same setting of Poisson arrivals, I consider the problem of choosing the optimal set of prices for multiple customer classes that are differentiated in their per-unit-time willingness to pay. I present key structural insights into the optimal prices in this case, finally culminating in a practically attractive iterative procedure that converges to these prices under standard regularity assumptions on the distribution of customer valuations. 

The analysis hinges on the observation that the per-unit-time opportunity cost of accepting a job from any customer class is in fact the optimal earning rate. If this rate is known, then the optimal pricing problem decomposes across classes into a set of independent optimal pricing decisions with a reserve price equal to the optimal earning rate. Effectively, these are basic optimal product pricing decisions. Since the optimal rate is not known, a natural iterative scheme for a worker is to start with a guess for the optimal  earning rate, compute the optimal prices independently for each customer class, estimate the resulting earning rate, recompute the optimal prices, and so on. I show that under the regularity assumptions on the distribution of valuations, this procedure indeed converges to the set of optimal prices and the optimal earning rate.

Finally, I discuss the issue of worker competition on these platforms. I show that with undifferentiated workers, an equilibrium vector of prices typically fails to exist. But for workers that are differentiated by their quality (as signaled by their ratings on these platforms, for instance), I propose a natural customer behavior model under which a price equilibrium exists, and can be efficiently computed. Under this model, that I term ``Best-I-can-afford" (BICA), each arriving customer chooses the highest quality available worker that she can afford. Under this model, I show that the workers can hierarchically solve monopolistic optimal pricing problems to arrive at the price equilibrium. This in particular implies that all the structural insights into the optimal pricing problem under the monopolistic setting carry over to the competitive setting from the perspective of any single worker.

The paper is organized as follows. I discuss relevant literature in the remainder of this section. In Section~\ref{sec:homo}, I present the model and the result on the optimality of a single hourly price with statistically identical customers and Poisson arrivals. In Section~\ref{sec:brittle}, I discuss examples of certain related settings where this prior result doesn't hold. In Section~\ref{sec:hetero}, I consider the case with multiple customer classes, where I analyze the structure of the optimal prices and finally present the iterative algorithm for computing these prices. In Section~\ref{sec:competition}, I consider the issue of worker competition. In Section~\ref{sec:extapp}, I end the paper with a discussion of possible extensions of the formulation, its applicability to practical settings, and possible future directions. 

\subsection{Related work}
Platform pricing in two-sided markets has received considerable attention in literature. \cite{rochet2003platform} pioneered the study of monopolistic pricing in such platforms; see also~\cite{rochet2006two}. This analysis was later generalized and extended to multi-sided platforms in \cite{weyl2010price}. \cite{caillaud2003chicken} and \cite{armstrong2006competition} additionally consider competition across platforms. These works abstract away the operational details of interactions on these platforms and exogenously specify how utilities of entities depend on the number of entities on the other side due to network effects. The focus is on traditional microeconomic questions like the impact of governance structures on the efficiency of market outcomes, whether and how prices internalize network effects, the impact of the agents' ability to multi-home, the impact of subscription based vs. transaction based pricing, etc. \cite{bakos2008design} considers the impact of investments into modulating the strength of network effects along with pricing decisions. Later works analyze the design of pricing and fee structures in two-sided market platforms while explicitly modeling their impact on the buyer/seller incentives and the resulting competitive equilibrium prices for trading that emerge on the platform.  Works in this spirit include \cite{economides2006two}, \cite{lin2011innovation}, and more recently, \cite{birge2018optimal}. These works assume a large market model with a continuum of entities transacting, due to which, a) price-taking behavior and competitive equilibrium emerge as natural determinants of trading prices, and b) distinction between products and services disappears.

In contrast to the platform perspective espoused in these works, the present paper focuses instead on the optimal pricing problem from the perspective of workers operating on labor platforms, while assuming the fees charged by the platforms as fixed. Moreover, in modeling competition, I assume a fixed and finite set of workers displaying fully strategic behavior in setting prices as opposed to assuming that workers are price-taking and their supply can be modulated by market prices. Because of this, the randomness inherent to the service aspect of the transactions cannot be neglected in the model, and in fact, it plays a crucial role in the analysis. 

As mentioned in the introduction, the decentralized nature of these pricing decisions for trades on the platform is different from the centralized pricing seen in ride-sharing platforms and analyzed in several recent works; c.f. \cite{banerjee2015,bimpikis2019spatial,cachon2017role,castillo2017surge}. Another important distinction in these works is the spatial nature of demand and supply, which is not modeled in the present work.

%Several works analyze the design of pricing and fee structures in two-sided market platforms while explicitly modeling the impact on the buyer/seller incentives and the resulting competitive equilibrium prices emerging on the platform. Works in this spirit include \cite{Ozan, IS}. These works assume a large market model with a continuum of entities transacting, due to which a) price-taking behavior and competitive equilibrium emerges as natural determinants of prices, and b) distinction between products and services disappears. The present work, on the other hand, focuses on the pricing decisions faced by a finite number of competing entities.

The present work is closely related to the literature on pricing in service systems; see \cite{hassin2003queue} for a survey. Three works most relevant to the present setting are \cite{ziya2006optimal}, \cite{ziya2008note} and \cite{caro2012optimal}. \cite{ziya2006optimal} considers the optimal pricing problem faced by a service facility with a single server and a queue with a finite capacity (which could be 0, as in our setting), when the customer valuations are drawn from a common distribution. Both \cite{ziya2008note} and \cite{caro2012optimal} analyze the case of multiple heterogeneously distributed customer classes in a similar setting. Although these settings are more general, the more focused analysis of the setting without queueing in the present work, motivated by on-demand platforms, results in several novel results and insights that do not hold in the general setting. In particular, the non-discrimination result with identically distributed valuations across customer classes holds only in this setting (as shown in the example in Section~\ref{sec:patient}). Similarly, the efficient procedure for computing optimal prices with heterogeneous customers doesn't necessarily converge in the presence of queues as I show in Remark~\ref{rem:queue} at the end of Section~\ref{sec:hetero}. Additionally, these works do not consider the issue of price competition amongst servers, neither do they consider maximization of discounted earnings as an objective.

\section{Homogeneous customers}\label{sec:homo}
Consider a single worker on an on-demand service platform. Customers are of $K$ types. Let the set of types be denoted as $\cK\triangleq\{1,\ldots,K\}$. Service requests to the worker from customers of type $k$ arrive according to a Poisson process with the rate of $\lambda_k$ per hour. The durations of the jobs they bring are distributed according to $G_k$ with mean $1/\mu_k$ hours. Let $\rho_k\triangleq \lambda_k/\mu_k$ denote the \emph{load} of the customer channel $k$ and let $\rho\triangleq \sum_{k}\rho_k$ denote the total load. Each customer has a maximum price $v$ that she is willing to pay per hour, which is drawn independently from a common distribution $F$. Let $\overline{F} = 1-F$ denote the tail distribution function. The worker chooses an hourly rate $p_k$ for job type $k$. An arriving customer's job is accepted by the worker only if she is idle and if the customer is willing to pay the hourly rate. While the worker is busy working on the job, all the arriving customers are lost forever (they are assumed to have chosen some other worker on the platform or to have left the platform altogether). While the worker is busy, she accrues a cost of $c$ per unit time. The goal of the worker is to choose the prices $p_k$ that maximize her long-run rate of earning.

We first derive an expression for the long-run average earning of the worker as a function of the price vector $\bp$. Observe that since the arrival process is Poisson, the total earning until time $t$, denoted as $R(t)$, is a renewal reward process where each renewal cycle consists of the worker becoming idle after finishing a job, then accepting the first paying job that arrives (such jobs arrive at the rate $\overline{F}(p_k)\lambda_k$ for each type $k$), and then finishing the job. Let $W_1$ be the total earning in the first cycle and let $S_1$ be the length of that cycle. Then from the renewal reward theorem \citep{gallager2013stochastic}, the long-run average earning of the worker as a function of $\bp$ is given by:
\begin{align}
\cR(\bp) = \lim_{t\rightarrow\infty} \frac{R(t)}{t} = \frac{\textup{E}(W_1)}{\textup{E}(S_1)}\,\,w.p.\,1.
\end{align}
Now since the arrival process is Poisson, the expected time till the first paying job arrives in a renewal cycle is $1/(\sum_{k'\in\cK} \lambda_{k'}\overline{F}(p_{k'}))$. Also, the first paying job that arrives in a renewal cycle is of type $k$ with probability $\lambda_k\overline{F}(p_k)/(\sum_{k'\in\cK} \lambda_{k'}\overline{F}(p_{k'}))$. Thus we have
$$\textup{E}(W_1) =   \frac{\sum_{k\in\cK}\frac{\lambda_k}{\mu_k}\overline{F}(p_k)(p_k-c)}{\sum_{k'\in\cK} \lambda_{k'}\overline{F}(p_{k'})}$$
and 
$$\textup{E}(S_1) = \frac{1}{\sum_{k'\in\cK} \lambda_{k'}\overline{F}(p_{k'})} + \frac{\sum_{k\in\cK}\frac{\lambda_k}{\mu_k}\overline{F}(p_k)}{\sum_{k'\in\cK} \lambda_{k'}\overline{F}(p_{k'})}.$$
We finally get 
\begin{align}
\cR(\mathbf{p}) = \frac{\sum_{k\in\cK}\rho_k(p_k-c)\overline{F}(p_k)}{1+\sum_{k\in\cK} \rho_k\overline{F}(p_k)}.\label{revexp}
\end{align}
The following result is the first main finding of the paper.
\begin{theorem}\label{thm:main}
There exists an optimal strategy for the worker that chooses the same hourly price across all customer types.
\end{theorem}
\proof{Proof.}
Recall that $\rho =\sum_{k\in\cK} \rho_k$ and denote $\alpha_k = \rho_k/\rho$. Then for any price vector $\bp$, we have,
\begin{align*}
\cR(\mathbf{p}) = \frac{\rho\sum_{k\in\cK}\alpha_k(p_k-c)\overline{F}(p_k)}{\sum_{k\in\cK}\alpha_k(1+ \rho\overline{F}(p_k))}\stackrel{(a)}{\leq} \rho\max_k\frac{(p_k-c)\overline{F}(p_k)}{1+ \rho\overline{F}(p_k)}\leq \max_{p}\frac{\rho(p-c)\overline{F}(p)}{1+ \rho\overline{F}(p)}.
\end{align*}
But the latter is the problem of choosing the single optimal hourly price for all jobs, thus proving the result. %, given that the total load is $\rho =\sum_{k\in\cK} \rho_k$. 
Inequality (a) relies on the following argument. Let $\Delta = \{\bar{\alpha}\in\mathbb{R}^K; \sum_{k\in\cK}\alpha_k =1; \, \alpha_k\geq 0 \,\,\forall k\in\cK\}$. Suppose that  $\mathbf{a} = (a_1,a_2,\cdots,a_k)$ is a non-negative vector and $\mathbf{b} = (b_1,b_2,\cdots,b_k)$ is a positive vector. Define
\begin{align*}
\omega^* = \max_{\bar{\alpha}\in\Delta} \frac{\sum_{k\in\cK}\alpha_ka_k}{\sum_{k\in\cK}\alpha_kb_k},
\end{align*}
and let $\bar{\alpha}^*$ be the maximizer. Then for any $\bar{\alpha}\in\Delta$, we have
\begin{align*}
\omega^* \geq \frac{\sum_{k\in\cK}\alpha_ka_k}{\sum_{k\in\cK}\alpha_kb_k}.
\end{align*}
This implies that for any $\bar{\alpha}\in\Delta$,
\begin{align*}
 \sum_{k\in\cK}\alpha_k(a_k-w^*b_k) \leq 0.
\end{align*}
This inequality is an equality only when $\bar{\alpha} =\bar{\alpha}^*$. This implies that $\bar{\alpha}^*$ is the maximizer of $\sum_{k\in\cK}\alpha_k(a_k-w^*b_k)$. But this implies that $\alpha^*_k>0$ if and only if $a_k-w^*b_k = 0$, i.e., if $w^* = a_k/b_k$. This implies that $w^* = a_k/b_k$ for some $k\in\cK$, which implies that $w^* = \max_{k\in\cK} a_k/b_k$. 
%This implies inequality (a).
\halmos
\endproof

\subsection{Discounted earnings}
In this section, I show that discounting the earnings over time does not affect the previous result. As before, let $p_k$ be the hourly price for job type $k$. Let $X(t)\in \{0\}\cup\cK$ be the state of the worker at time $t$, beginning from the state $X(0) = 0$. Here, the state $0$ signifies that the worker is idle, and state $k\in\cK$ signifies that the worker is busy with a job of type $k$. Then $(X(t))_{t\in\mathbb{R}_{\geq0}}$ is a continuous time stochastic process that is c\`adl\`ag, i.e., its sample paths are right continuous with left limits. Let $R(x)$ be the earning rate in state $x$, where $R(0) = 0$ and $R(k) = p_k-c$ for $k\in\cK$. Let $\gamma>0$ be the discount factor. Then the total expected discounted reward is given by:
\begin{align}
\cR^{\gamma}(\bp) = \textup{E}[\int_{0}^\infty R(X(t))\me^{-\gamma t}dt\mid X(0) = 0].
\end{align}
Let $T$ be an exponential random variable with mean $1/\gamma$, independent of $(X(t))_{t\in\mathbb{R}_{\geq0}}$. Consider a modified stochastic process $(\hat{X}(t))_{t\in\mathbb{R}_{\geq0}}$, defined as $\hat{X}(t)= X(t)\mathbf{1}_{\{t< T\}} + a\mathbf{1}_{\{t\geq T\}}$ by introducing a new absorbing state $a\triangleq K+1$ such that $R(a) = 0$. Clearly, if $(X(t))_{t\in\mathbb{R}_{\geq0}}$ is c\`adl\`ag, then $(\hat{X}(t))_{t\in\mathbb{R}_{\geq0}}$ is c\`adl\`ag as well. Then it is straightforward to see that $\cR^{\gamma}(\bp)$ is the expected total earning in this modified process, i.e., 
\begin{align}
\cR^{\gamma}(\bp) = \textup{E}[\int_{0}^\infty R(\hat{X}(t))dt\mid \hat{X}(0) = 0].
\end{align}
Define $\beta(0) = \cR^{\gamma}(\bp)$, and for $k\in\cK$, define,
\begin{align}
\beta(k) = \textup{E}[\int_{u}^\infty R(\hat{X}(t))dt\mid \hat{X}(u) = k,  \hat{X}(u_-)= 0],
\end{align}
where $\hat{X}(u_-)= \lim_{t\uparrow u}\hat{X}(t)$. Thus, $\beta(k)$ is the expected total earning until absorption starting from the state where a job of type $k$ has just been accepted by the worker. For each $k\in\cK$, let $X_k\sim G_k$, and let $Y\sim\exp(1/\gamma)$, such that $Y$ is independent of $X_k$. We then get the following set of first step equations for computing $\beta(0)$. First, we have,
\begin{align}
\beta(0) = \frac{\sum_{k\in \cK}\lambda_k \overline{F}(p_k)\beta(k)}{\sum_{k\in \cK}\lambda_k\overline{F}(p_k) +\gamma},
\end{align}
where $\lambda_k \overline{F}(p_k)/(\sum_{k\in \cK}\lambda_k\overline{F}(p_k) +\gamma)$ is the probability that the process enters state $k$ before absorption. Also, for each $k\in\cK$, we have,
\begin{align}
\beta(k) = (p_k-c) \textup{E}[\min(X_k,Y)] +P(X_k<Y)\beta(0).
\end{align}
The first term in the expression on the right is the expected earning after accepting a job until either the job is finished or the process enters the absorbing state. The expected time till either of those two events occur is $\textup{E}[\min(X_k,Y)]$. The second term results from the fact that the process enters state $0$, i.e., the job gets finished before absorption, with probability $P(X_k<Y)$. Solving, we get,
\begin{align*}
\beta(0)(\sum_{k\in \cK}\lambda_k\overline{F}(p_k) +\gamma) = \sum_{k\in\cK}\lambda_k \overline{F}(p_k)(p_k-c)\textup{E}[\min(X_k,Y)] + \beta(0) \sum_{k\in\cK}\lambda_k \overline{F}(p_k)P(X_k<Y),
\end{align*}
or,
\begin{align*}
\beta(0)(\sum_{k\in \cK}\lambda_k\overline{F}(p_k)P(Y\leq X_k) +\gamma) = \sum_{k\in\cK}\lambda_k \overline{F}(p_k)(p_k-c)\textup{E}[\min(X_k,Y)].
\end{align*}
Thus, denoting $\hat{\rho}_k = \lambda_k\textup{E}[\min(X_k,Y)]$, we finally have,
\begin{align}
\beta(0)&= \cR^{\gamma}(\bp)= \frac{\sum_{k}\hat{\rho}_k(p_k-c)\overline{F}(p_k)}{\gamma+\sum_{k\in\cK} \hat{\rho}_k\overline{F}(p_k)\frac{P(Y\leq X_k)}{\textup{E}[\min(X_k,Y)]}}.
\end{align}
We now need the following fact.
\begin{lemma}
Let $X$ and $Y$ be independent non-negative random variables such that $Y$ is exponentially distributed with mean $1/\gamma$. Then,
$$\frac{P(Y\leq X)}{\textup{E}[\min(X,Y)]} = \gamma.$$
\end{lemma}
\proof{Proof.}
We have,
\begin{align}
\textup{E}[\min(X,Y)\mid X] = X\exp(-\gamma X) +(1-\exp(-\gamma X) )\textup{E}(Y\mid Y\leq X).
\end{align}
Conditioned on the event $\{Y\leq X\}$, $Y$ has a truncated exponential distribution, and its mean can be computed to be (see Chap. 4, Lemma 4.3 in \cite{olive2008applied}),
\begin{align}
\textup{E}(Y\mid Y\leq X) = \frac{1}{\gamma}\bigg(\frac{1-(X\gamma+1)\exp(-X\gamma)}{1-\exp(-X\gamma)}\bigg).
\end{align}
Thus,
\begin{align}
\textup{E}[\min(X,Y)\mid X] = \frac{1}{\gamma}(1-\exp(-\gamma X)).\label{expo}
\end{align} 
This, coupled with the fact that $P(Y\leq X\mid X) = 1-\exp(-\gamma X)$ implies the result. \halmos
\endproof
Thus, we finally have,
\begin{align}
\cR^{\gamma}(\bp) &= \frac{\sum_{k}\hat{\rho}_k(p_k-c)\overline{F}(p_k)}{\gamma(1+\sum_{k\in\cK} \hat{\rho}_k\overline{F}(p_k))}.
\end{align}
Using the same argument as that in the proof of Theorem~\ref{thm:main}, it is straightforward to establish that there exists an optimal policy that sets the same hourly price for each customer class. This is the policy that maximizes the long-run average earning given that the total load is $\hat\rho = \sum_{k\in\cK}\hat\rho_k$, where $\hat{\rho}_k = \lambda_k\textup{E}[\min(X_k,Y)]$. For instance, if the job durations are exponentially distributed, then $\hat{\rho}_k = \lambda_k/(\mu_k +\gamma)$.

\section{Related settings}\label{sec:brittle}
In this section, I take a short detour and present examples to show that Poisson arrivals are not sufficient for the previous result in an arbitrary system. 
\subsection{Worker specific queues}\label{sec:patient}
Consider the case where the customers could be patient, i.e., they are willing to wait for the worker to become free before choosing someone else. This could naturally model traditional freelancing settings where the worker has developed a sustained relationship with a class of customers who prefer her services over other workers. I numerically demonstrate than in these cases, the previous result may not hold, i.e., even under Poisson arrivals, price discrimination may be necessary for optimality despite identically distributed customer preferences. 
%The reason is that the per-unit-time opportunity costs for shorter jobs are in general smaller than those for longer jobs. 

%The intuition is as follows. Suppose that the hourly willingness to pay is identically distributed for all customers and every customer that arrives waits for 1 hour on an average before choosing someone else. Suppose that at any time, at most one person can be in the queue (assume that customers leave if they see a customer already waiting). There are two classes of customers: class A customers bring a job of duration 5 mins and class B customers bring a job of duration 2 hours. Now, accepting a job from class A incurs negligible opportunity cost relative to its duration, whereas accepting a class B job results in a relatively higher opportunity cost relative to its duration. This is because in the first case the worker is not expected to lose any waiting customers, whereas in the latter situation some waiting customers will be lost with a high probability. Thus in this case the total expected opportunity cost does not grow linearly in the duration of the job. And thus the optimal hourly prices for the shorter jobs are expected to be lower.  I numerically verify that this is indeed the case. 

Suppose that there are two classes of customers, A and B. The hourly willingness to pay of each customer is uniformly distributed in $[0,1]$. Customers from class A arrive at the rate of $\lambda_A$ customers per hour and those from B arrive at the rate of $\lambda_B$ customers per hour. The job durations of both classes are exponentially distributed: class A with mean $1/\mu_A$ hours and class B with mean $1/\mu_B$ hours. Suppose that arriving customers are willing to wait indefinitely if the worker is busy and if they are next in line. If an arriving customer sees another customer waiting, then she leaves the system. In other words, the queue has capacity $1$ and customers do not leave without being served once they are in queue.\footnote{If the queue has infinite capacity and customers wait indefinitely, then there are no externalities imposed by accepting any job and no price discrimination is necessary if the customers' per-unit-time willingness to pay is identically distributed.}   

\begin{figure}[htb]
\begin{center}
\includegraphics[width=3in,angle=0]{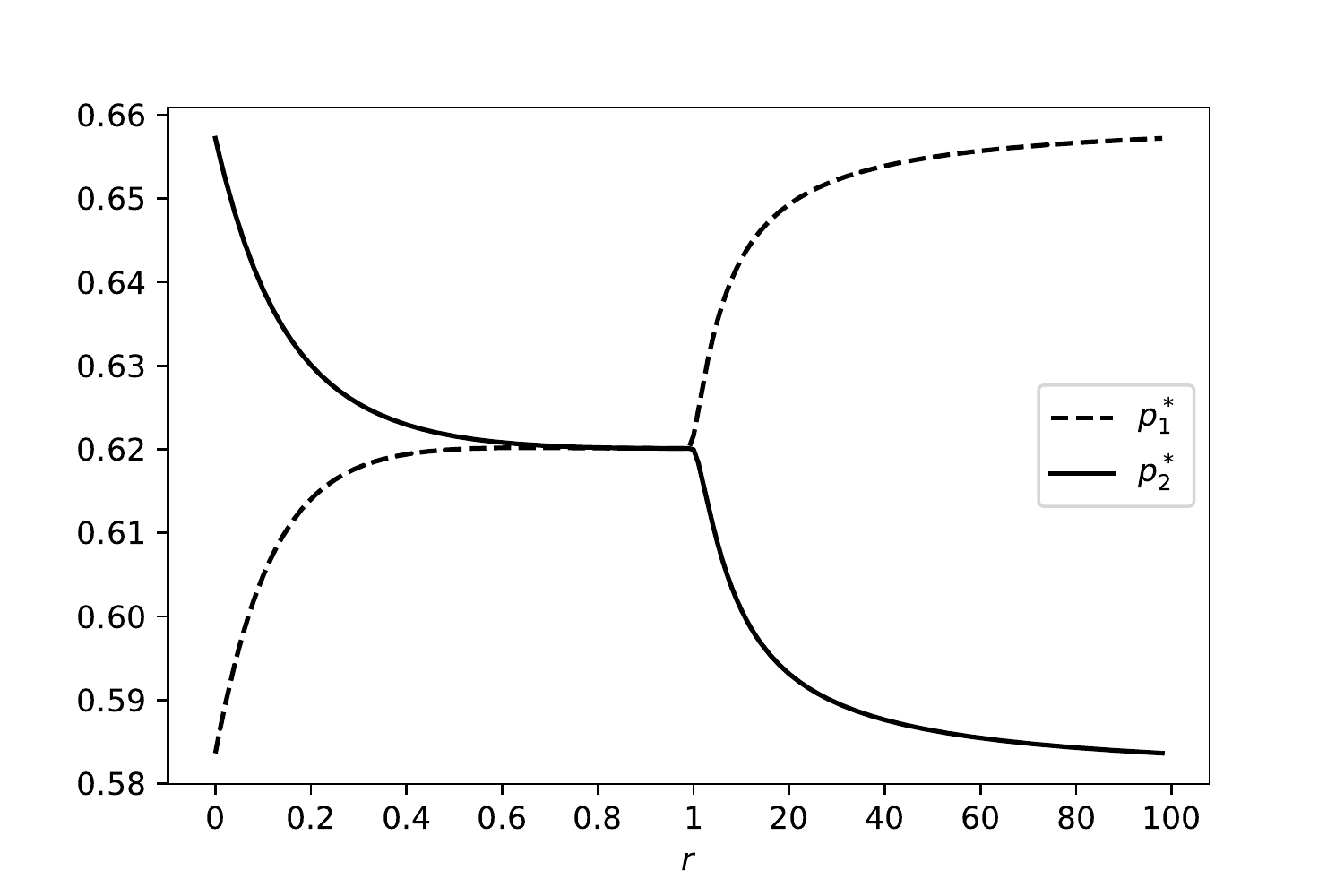}
\caption{Optimal prices as a function of $r$.}
\label{fig:patient}
\end{center}
\end{figure}

For a pair of prices $p_A$ and $p_B$ for the two classes, since the arrival process is Poisson, the reward process till any time $t$ is a renewal reward process. Each renewal cycle begins with the worker being idle, and ends when the worker finishes a job and there is no customer in queue. The long-run average rate of earning in this case can be computed to be,
\begin{align}
\mathcal{R}(p_A,p_B) &= \frac{\rho_Ap_A(1-p_A) + \rho_Bp_B(1-p_B)}{\frac{\lambda_A\mu_A(1-p_A)+\lambda_B\mu_B(1-p_B) +\mu_A\mu_B}{(\lambda_A(1-p_A)+\lambda_B(1-p_B) +\mu_A)(\lambda_A(1-p_A)+\lambda_B(1-p_B) +\mu_B)} + \rho_A(1-p_A) +\rho_B(1-p_B)}.\label{hanger}
\end{align}
The details of this computation can be found in the Appendix. Suppose that we fix $\lambda_A = \mu_A = 1$ and set $\lambda_B = \mu_B = r$, so that $\rho_A=\rho_B=1$. In Figure~\ref{fig:patient}, I plot the optimal prices as $r$ varies from $0$ to $1$ and then from $1$ to $100$.\footnote{The optimal prices were computed using the scipy.optimize package in Python.}  This allows us to see the effect on the optimal prices of changing the job length of class $B$ while keeping the overall load  the same as that of class $A$. Observe that when the expected job length of class $B$ is larger than that of class $A$ ($r<1$), the optimal price $p^*_B$ is larger than $p^*_A$. The inequality is reversed when $r>1$, i.e., when class $B$ jobs are shorter than class $A$ jobs. This suggests that in these cases, the expected opportunity cost doesn't necessarily grow linearly in the length of the job. Intuitively, when the worker starts working on a job, the rejections of incoming jobs only begin when a new job arrives in the queue. Thus smaller jobs impose relatively smaller externalities per unit time since there is a greater chance that they finish before a new job arrives in the queue.   

%\subsection{Cost for lost customers}
%The worker may wish to incorporate a cost for every lost customer who is willing to pay the price posted by the worker. In this case, one can again show that with Poisson arrivals, price discrimination across different customer classes with identically distributed hourly willingness to pay is unnecessary, even if the cost associated with a lost customer is different across the different classes. Formally, suppose that $r_k$ is the cost per customer lost of class $k$. Then from a calculation similar to one in deriving equation~\eqref{revexp}, for a vector of prices $\mathbf{p}$, the net earning rate of the worker is, 
%\begin{align}
%\cR(\mathbf{p}) = \frac{\sum_{k\in\cK}\rho_k(p_k-c-\sum_{k\in\cK}\lambda_kr_k \overline{F}(p_k))\overline{F}(p_k)}{1+\sum_{k\in\cK} \rho_k\overline{F}(p_k)}.\label{revexp}
%\end{align}
%The term $\sum_{k\in\cK}\lambda_kr_k$
\subsection{A hybrid setting}
%In a true on-demand platform, the worker is expected to take up a customer immediately on arrival and finish the job completely before she can accept another customer. This setting is the main focus of the present paper.  
On a related note to the setting described in the previous section, in platforms where there is a mixture of on-demand and patient customer classes, price discrimination across these classes is typically necessary even if the distribution of the customers' hourly willingness to pay and arrival/job-duration characteristics are identical. To see this, suppose that there are two classes of customers: On-demand (A) and Patient (B). On-demand customers have to be served immediately, and Patient customers are willing to wait as long as the average waiting time is finite (i.e., the queue doesn't explode). Moreover, preemption of a Patient job is allowed. Suppose that all job durations are exponentially distributed. The arrival rate and service rate of the On-demand class are $\lambda_A = \mu_A =1$. The arrival rate and service rate of the Patient class are $\lambda_B = 1/\sqrt{2} - \epsilon$ and  $\mu_B =1$ for some small $\epsilon>0$ (the choice may seem mysterious but will soon become clear). Suppose that the hourly valuations of the customers are distributed uniformly in $[0,1]$.

Now clearly, as long as the effective residual service rate available for the Patient class due to time spent in serving the On-demand class is larger than $\lambda_B$, the Patient class imposes no externality on the system, since accepting a Patient customer incurs no loss in revenue to the worker. The optimal hourly price for the Patient class can then be computed to be $0.5$. The question is whether the effective service rate for the Patient class is sufficient. Now, turning to the On-demand class, it is clear that accepting a job imposes an externality on the system. If this was the only class on the platform, then the optimal price is the one that maximizes,
$$\frac{x\overline{F}(x)}{1+\overline{F}(x)} = \frac{x(1-x)}{2-x}.$$
The optimal price can be computed to be $p^*_A = 2-\sqrt{2}$. At this price, the long-run fraction of time that the worker is not working on an On-demand job can be computed to be $1/(2-p^*_A)=1/\sqrt{2}$ (this is simply the ratio of the expected time until arrival of an On-demand job in a renewal cycle and the expected total length of a renewal cycle). Hence, the effective service rate available for the Patient customers is $1/\sqrt{2}\times 1$ which is more than $\lambda_B$. Thus we conclude that the optimal hourly price for the On-demand customers is $2-\sqrt{2}$ and that for the Patient customers is $0.5$.

This example suggests that in a general setup where there exist different classes of customers that are differentiated in their delay tolerance, which can be modeled by their job deadlines, price discrimination may be necessary for optimality despite identically distributed preferences.

%If the worker can simultaneously work on multiple jobs at the same time without affecting the service times of the jobs, then again the result doesn't hold: price discrimination based on job lengths may be necessary. Consider a setting with 2 customer classes, A and B, each arriving with rate $\lambda$, with class $i$ service times exponentially distributed with mean $1/\mu_i$ for $i=1,\,2$. In this case, again the reward process is a renewal reward process. Each renewal cycle begins with an idle time of mean length $1/2\lambda$. Suppose that the hourly willingness to pay of both customer types is uniformly distributed in $[0,1]$. Let $P_0$ denote the mean reward earned in each renewal cycle. Let $P_1$ $(P_2)$ denote the mean reward earned after a customer of class $1$ (class $2$) is accepted in an empty system, until the system becomes empty again. Let $T_0$ denote the mean duration of each renewal cycle. Let $T_1$ $(T_2)$ denote the mean duration after a customer of class $1$ (class $2$) is accepted in an empty system, until the system becomes empty again. Let $p_1$ and $p_2$ be the prices set for the two customer classes. Let $\lambda'_i \triangleq \lambda(1-p_i)$ for $i=1,2$. We then have the following first-step equations:

%\begin{align}
%P_0 &= \frac{\lambda'_1P_1}{\lambda'_1+\lambda'_2} +\frac{\lambda'_2P_2}{\lambda'_1+\lambda'_2}\\
%P_1 &= 
%\end{align}
\subsection{Non-exponential discounting}
One can show that price discrimination across customer classes because of differing arrival/duration characteristics may be necessary despite Poisson arrivals if the time discounting is non-exponential. We saw earlier that exponential discounting is equivalent to an exponentially distributed time horizon. Suppose instead that the time horizon is exponentially distributed with rate $1$ with probability $1/2$ and with rate $2$ with probability $1/2$. Suppose there are 2 classes of customers, A and B. The arrival rates are $\lambda_A = \lambda_B = 1$. The job durations are exponentially distributed with rates $\mu_A = 1$ and $\mu_B = 2$.  Suppose that the hourly prices are $p_A$ and $p_B$ for the two classes. In the event that the time horizon is exponentially distributed with rate $\gamma$, the expected earning can be computed to be,
\begin{align}
\frac{\hat{\rho}_Ap_A(1-p_A)+\hat{\rho}_Bp_B(1-p_B)}{1+\hat{\rho}_A(1-p_A)+\hat{\rho}_B(1-p_B)},
\end{align}
where $\hat{\rho}_A = \lambda_A/(\mu_A+\gamma)$ and $\hat{\rho}_B = \lambda_B/(\mu_B+\gamma)$. Thus the expected earnings can be computed to be,
\begin{align}
\cR(p_A, p_B) &= \frac{1}{2} \times \frac{(1/2) p_A(1-p_A)+ (1/3)p_B(1-p_B)}{1+(1/2) (1-p_A)+(1/3) (1-p_B)}\nonumber \\
&~~+ \frac{1}{2} \times \frac{(1/3) p_A(1-p_A)+ (1/4)p_B(1-p_B)}{1+(1/3) (1-p_A)+(1/4) (1-p_B)}
\end{align}
The optimal prices can be computed to be $p^*_A \approx 0.56761$ and $p^*_B \approx 0.567089$.\footnote{Optimal prices computed using the scipy.optimize package in Python} Thus in this case, price discrimination becomes necessary for optimality.

\section{Price differentiation with heterogeneous customers}\label{sec:hetero}
Coming back to the on-demand setting where there are no per-worker queues, it is now clear that under Poisson arrivals, any price differentiation must stem from the fact that the beliefs about the customers' willingness to pay are different across different classes. In this section, I present structural insights into the optimal prices in this case and finally present a procedure to compute these prices under certain standard assumptions. Let $F_k$ be the distribution of hourly valuations of customers in class $k\in\cK$, with support on $[0,\bar{v}_k]$. I assume that each $F_k$ is differentiable on its support (with right and left derivatives defined at $0$ and $\bar{v}_k$, respectively). Let $\bar{F}_k = 1-F_k$ and $f_k$ denote the corresponding tail distribution functions and densities, respectively. It is straightforward to establish as we did earlier (i.e., the derivation of \eqref{revexp}), that for a price vector $\bp$, the long-run average earning is given by:
\begin{align}
\cR(\mathbf{p}) = \frac{\sum_{k}\rho_k(p_k-c)\overline{F}_k(p_k)}{1+\sum_{k\in\cK} \rho_k\overline{F}_k(p_k)}.
\end{align}
The problem of maximizing the long-run rate of earning is thus defined as:
\begin{align}
\max_{\bp: p_k\in [0,\bar{v}_k]\, \forall k}\cR(\mathbf{p}).\label{opt2}
\end{align}
Let the optimal rate of earning be denoted by $\cR^*$. Then for any feasible set of prices $\bp$, we have that
\begin{align}
&\cR^* \geq   \frac{\sum_{k}\rho_k(p_k-c)\overline{F}_k(p_k)}{1+\sum_{k\in\cK} \rho_k\overline{F}_k(p_k)}\nonumber\\
&\Rightarrow \cR^*(1+\sum_{k\in\cK} \rho_k\overline{F}_k(p_k)) \geq  \sum_{k}\rho_k(p_k-c)\overline{F}_k(p_k)\nonumber\\
&\Rightarrow  \cR^* \geq  \sum_{k}\rho_k(p_k-c-\cR^*)\overline{F}_k(p_k).\label{arg}
\end{align}
This expression is an equality if and only if $\bp$ is the vector of optimal prices $\bp^*$ (which need not be unique). This also implies that the optimal prices are the maximizers of the function  $\sum_{k}\rho_k(p_k-c-\cR^*)\overline{F}_k(p_k)$. Since this function is separable across the classes, this in turn implies that any optimal price for class $k$, $p^*_k$, maximizes  $\overline{F}_k(p_k)(p_k-c-\cR^*)$. This fact has an intuitive interpretation. The optimal earning rate $\cR^*$ is indeed the opportunity cost per hour of being busy. Thus, given $\cR^*$, one simply solves the optimal pricing problem for each class assuming that the total hourly cost for serving that class is $c+ \cR^*$. This is a standard optimal pricing problem in mechanism design; c.f.  \cite{myerson1981optimal}.

%The optimality conditions imply that given the optimal opportunity costs, the optimal pricing problem can be decomposed across the different customer classes. This decomposition is very attractive from a practical standpoint. 
For a fixed hourly opportunity cost $\cR\geq 0$, for each $k\in\cK$, consider the optimization problem, 
\begin{align}
\max_{p\in[0,\bar{v}_k]}\overline{F}_k(p)(p-c-\cR).\label{mid}
\end{align}
The first derivative of the objective function is given by,
\begin{align}
\overline{F}_k(p)-(p-c-\cR)f_k(p) = -f_k(p)\bigg(p-c-\cR-\frac{\overline{F}_k(p)}{f_k(p)}\bigg).\label{derivative}
\end{align}
%Then, from the Karush-Kuhn-Tucker (KKT) optimality conditions, we obtain that $p_k(\cR)$ satisfies:
%\begin{align}
%p_k(\cR)  &=c+\cR + \frac{\overline{F}_k(p_k(\cR))}{f_k(p_k(\cR))}, &\textrm{if }p_k(\cR)< \bar{v}_k,\\
%p_k(\cR) &\leq c+ \cR + \frac{\overline{F}_k(p_k(\cR))}{f_k(p_k(\cR))},\, \textrm{i.e., } \bar{v}_k\leq c+\cR,&\textrm{if }p_k(\cR)= \bar{v}_k,
%\end{align}
%for all $i\in\cK$. 
Now suppose that the function $p-\overline{F}_k(p)/f_k(p)$ is strictly increasing in $p\in[0,\bar{v}_k]$ for each $k\in\cK$; if this holds we say that each distribution $F$ is \emph{strictly regular} (if $p-\overline{F}_k(p)/f_k(p)$ is non-decreasing then such a distribution is called \emph{regular} in mechanism design literature; c.f. \cite{myerson1981optimal}). For instance, this would hold if $F_k$ has a non-decreasing hazard rate for each $k\in\cK$, i.e., $f_k(p)/\overline{F}_k(p)$ is non-decreasing (this has been referred to as the \emph{Monotone Hazard Rate} or \emph{Increasing Failure Rate} assumption in the mechanism design literature). 

If the distributions are strictly regular, the entire function in the parenthesis in equation \eqref{derivative} is increasing. Thus, from the first order optimality condition, we can conclude that the optimal value of $p$ that solves \eqref{mid} is unique: it is $\bar{v}_k$ if $\bar{v}_k \leq c+\cR$, otherwise it is the unique value of $p$ that satisfies $p=c+\cR+\overline{F}_k(p)/f_k(p)$. With some abuse of notation, let us denote this optimal value as $p_k(\cR)$. It is straightforward to verify that the function $p_k(\cR)$ defined on $\mathbb{R}^+$ is continuous, and moreover differentiable at all points except $\bar{v}_k-c$. Also, both left and right derivatives exist at $\bar{v}_k-c$. Further, $p'_k(\cR)\geq 0$ for all $\cR<\bar{v}_k-c$ (the optimal price is non-decreasing in $\cR$) and  $p'_k(\cR)=0$ for all $\cR>\bar{v}_k-c$. At $\bar{v}_k-c$, we have the left derivative $p'_{k-}(\bar{v}_k-c)\geq 0$ and the right derivative $p'_{k+}(\bar{v}_k-c)= 0$.

Thus, under the strict regularity assumption, we know that the optimal prices $\bp^*$ are unique and they are $(p_k(\cR^*))_{k\in\cK}$. The optimal earning rate $\cR^*$ is of course, unknown. But we know that $\cR^*$ satisfies the fixed point relation:
\begin{align}
\cR^* = \frac{\sum_{k\in\cK}(p_k(\cR^*)-c)\rho_k\overline{F}_k(p_k(\cR^*))}{1+\sum_{k\in\cK} \rho_k\overline{F}_k(p_k(\cR^*))}.\label{fixed}
\end{align}
%A similar characterization has been obtained for a more general setting with a finite capacity queue in \cite{caro2012optimal}, where the uniqueness of the fixed point has also been claimed (without proof) under similar assumptions. 
The following result shows that given the regularity assumption, 1) this fixed point is unique, and 2) it is the maximizer of the response function. These two facts are key to obtaining a natural procedure for iteratively computing $\cR^*$.
\begin{lemma}
Suppose that the distributions $F_k$ are strictly regular for each $k\in\cK$. Then, there is a unique $\cR^*$ that satisfies the fixed point relation~\eqref{fixed}. Moreover, $\cR^*$ maximizes the function,
\begin{align}
M(\cR) =\frac{\sum_{k\in\cK}(p_k(\cR)-c)\rho_k\overline{F}_k(p_k(\cR))}{1+\sum_{k\in\cK} \rho_k\overline{F}_k(p_k(\cR))}.
\end{align}
\end{lemma}
\proof{Proof.}
First, note that a fixed point always exists since the optimal earnings and the optimal set of prices define such a fixed point. We first establish that it is unique. 
It is straightforward to verify that the function $M(\cR)$ defined on $\mathbb{R}^+$ is continuous, and differentiable at all points except $\{\bar{v}_k-c;\, k\in\cK\}$. Also at these points, both left and right derivatives exist (these properties follow from the corresponding properties of $p_k(\cR)$). Then at any point $\cR$ the right-derivative of $M(\cR)$ can be expressed as,
\begin{align}
M'_{+}(\cR) &=\sum_{k\in\cK}\frac{\partial M}{\partial p_k}p'_{k+}(\cR)\\
&=\sum_{k\in\cK} \frac{-\rho_kf_k(p_k(\cR))}{1+\sum_{k'\in\cK} \rho_{k'}\overline{F}_{k'}(p_{k'}(\cR))}\bigg(p_k(\cR)-c-M(\cR)-\frac{\overline{F}_k(p_k(\cR))}{f_k(p_k(\cR))}\bigg)p'_{k+}(\cR).
\end{align}
Here $p'_{k+}(\cR)$ is the right-derivative of $p_k$ at $\cR$. Let $\mathcal{R^*}$ be a fixed point of $M$, i.e., $M(\cR^*)=\cR^*$. Then we have,
\begin{align}
M'_{+}(\cR^*) &=\sum_{k\in\cK} \frac{-\rho_kf_k(p_k(\cR^*))}{1+\sum_{k'\in\cK} \rho_{k'}\overline{F}_{k'}(p_{k'}(\cR^*))}\bigg(p_k(\cR^*)-c-\cR^*-\frac{\overline{F}_k(p_k(\cR^*))}{f_k(p_k(\cR^*))}\bigg)p'_{k+}(\cR^*).
\end{align}
From the optimality conditions for $p_k(\cR^*)$, we have that $p_k(\cR^*)-c-\cR^*-\overline{F}_k(p_k(\cR^*))/f_k(p_k(\cR^*))\leq 0$. Moreover, if  $p_k(\cR^*)-c-\cR^*-\overline{F}_k(p_k(\cR^*))/f_k(p_k(\cR^*))< 0$ then $p'_{k+}(\cR^*) = 0$. Thus $M'_{+}(\cR^*)= 0$ for any $\cR^*$ such that $M(\cR^*)=\cR^*$. 

This implies that for any $\cR^*$, there is an $\epsilon$ such that for all $\cR$ that satisfy $\cR^*<\cR<\cR^*+\epsilon$, $M(\cR)<\cR$. For such an $\cR$, if $p_k(\cR)-c-\cR-\overline{F}_k(p_k(\cR))/f_k(p_k(\cR))= 0$, then $p_k(\cR)-c-M(\cR)-\overline{F}_k(p_k(\cR))f_k(p_k(\cR))>0$ and $p'_{k+}(\cR)\geq 0$, while if $p_k(\cR)-c-\cR-\overline{F}_k(p_k(\cR))/f_k(p_k(\cR))< 0$, then $p'_{k+}(\cR)= 0$. Thus, $M'_{+}(\cR)\leq 0$. This implies that the function $M(\cR)$ is non-increasing after $\cR^*$. Hence, there can be only one fixed point $\cR^*$. 

%Next, suppose that $\mathcal{R^*}$ is a fixed point of $M(\cR)$ and $M(\cR)$ is not differentiable at $\cR^*$. Then for all classes $k$ such that $\bar{v}_k =\cR^*$, we have the right-derivative of $p_k(\cR)$, equal to $0$. While for all other classes, we have $p_k(\cR^*)-c-\cR^*-\overline{F}_k(p_k(\cR^*))/f_k(p_k(\cR^*))= 0$. Thus, the right derivative of $M(\cR)$ equals $0$ for all $\cR^*$ such that $M(\cR^*)=\cR^*$. 

Next, we show that $\cR^*$ maximizes $M(\cR)$. This is straightforward, since for any $\cR$, $M(\cR)$ is an achievable revenue rate, achieved by simply using the prices $(p_k(\cR))_{k\in\cK}$. Hence, $M(\cR)\leq \cR^*=M(\cR^*)$.\halmos
%Now consider a point $\cR<\cR^*$. Since $M(0)>0$, and $M$ is continuous, we have $M(\cR)>\cR$. Now since $p_k(\cR)-c-\cR-\overline{F}_k(p_k(\cR))/f_k(p_k(\cR))\leq 0$, we have $p_k(\cR)-c-M(\cR)-\overline{F}_k(p_k(\cR))/f_k(p_k(\cR))<0$. On the other hand, $p'_{k+}(\cR)\geq 0$. Thus, $M'_{+}(\cR)\geq 0$ for $\cR< \cR^*$. This implies that the function $M$ is non-decreasing for $\cR<\cR^*$. This shows that $M(\cR)$ is maximized at $\cR^*$ (with a maximum value $\cR^*$). 
\endproof
We can now show the following main result of this section.
\begin{theorem}
Suppose that the distributions $F_k$ are strictly regular for each $k\in\cK$. Then starting from any $\cR_0\geq 0$, the sequence $(\cR_t)_{t\in\mathbb{N}}$ obtained by the relation $\cR_{t+1}=M(\cR_t)$ converges to the unique fixed point $\cR^*$ of $M(\cR)$. $\cR^*$ is the optimal rate of earning and the corresponding prices $p_k(\cR^*)$ for $k\in\cK$ are the optimal prices. 
\end{theorem}
\proof{Proof.}
Since $M(\cR)\leq \cR^*$, for any $\cR_0$, we have that $\cR_t\leq \cR^*$ for $t\geq 1$. Moreover, $\cR_{t+1}= M(\cR_t)> \cR_t$ for any $\cR_t<\cR^*$. Thus $(\cR_t)_{t\in\mathbb{N};\,t\geq 1}$ is a monotonically increasing sequence bounded by $\cR^*$, and hence converges to some $\cR'$ by the monotone convergence theorem. It must be that $\cR'=\cR^*$, since if not, then $\cR'<\cR^*$ and hence $M(\cR')>\cR'$, which contradicts the fact that $\cR'$ is the limit point. \halmos
\endproof

{\bf Example.} Consider two customer classes with loads $\rho_1=\rho_2=1$. The hourly willingness to pay of class $1$ is uniformly distributed in $[0,1]$, and that of class $2$ is uniformly distributed in $[0,2]$. Thus $\overline{F}_1(x) = (1-x)\mathbbm{1}_{\cR\in [0,1]} $ and $\overline{F}_2(x) = (1-x/2)\mathbbm{1}_{\cR\in [0,2]} $. Suppose that the hourly cost of service is $0$.  For a given hourly reserve price $\cR$, $p_k(\cR)$ solves $\max(p-\cR)(1-p)$ for $k=1$ and $\max(p-\cR)(1-p/2)$ for $k=2$. We thus obtain $p_1(\cR) = \frac{1+\cR}{2}\mathbbm{1}_{\cR\in [0,1]} + \mathbbm{1}_{\cR>1}$ and $p_2(\cR)= \frac{2+\cR}{2}\mathbbm{1}_{\cR\in [0,2]} + 2\mathbbm{1}_{\cR>2}$.

\begin{figure}[h]
\centering
\begin{minipage}[t]{.4\textwidth}
\centering
\includegraphics[width=2.8in,angle=0]{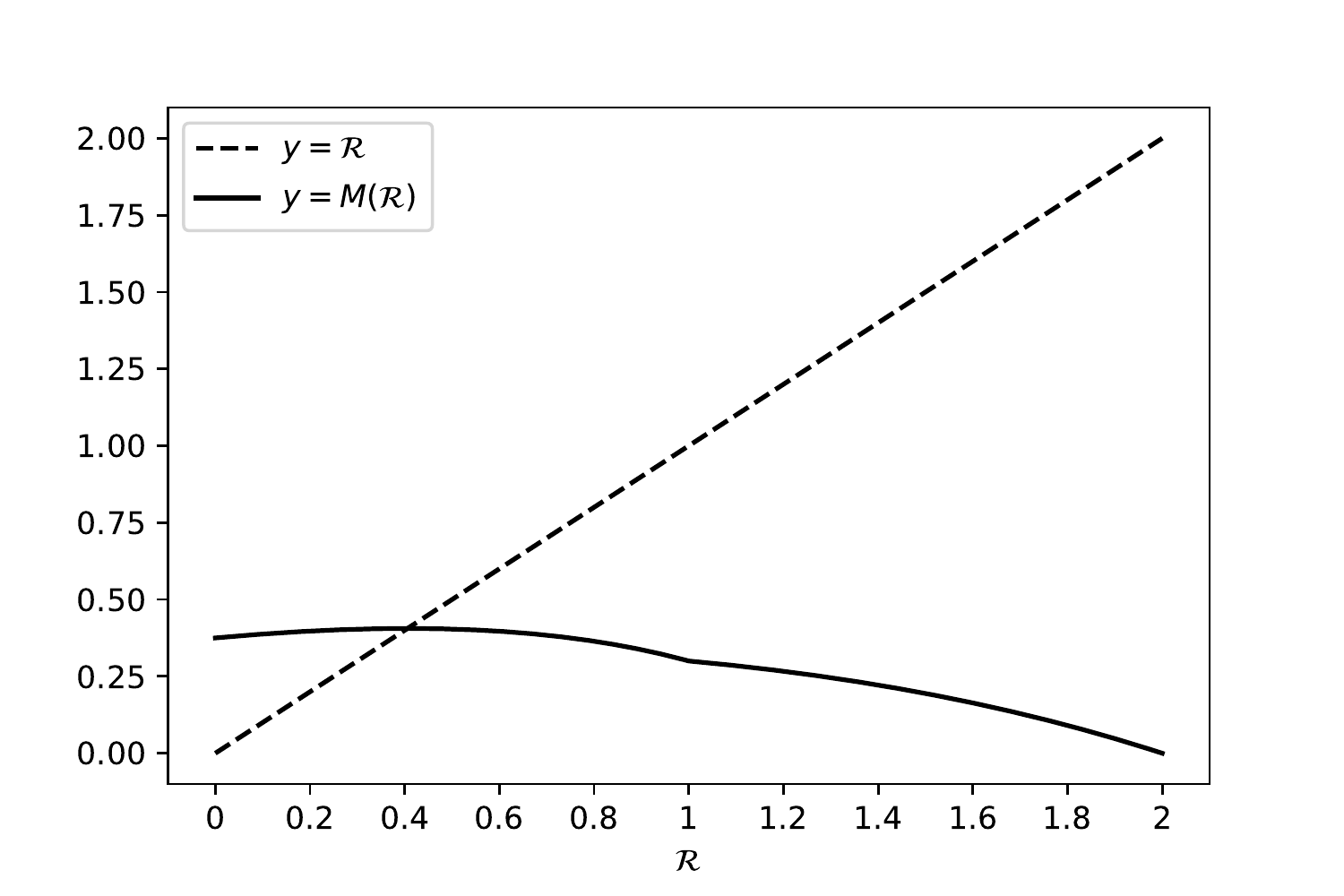}
\caption{The plot of $y=M(\cR)$ and the line $y=\cR$. The fixed point of $M(\cR)$ ($\cR^* \approx 0.40589$) is also the maximizer of $M(\cR)$.}
\label{fig:example}
\end{minipage}\hfill
\begin{minipage}[t]{.5\textwidth}
\centering
\includegraphics[width=2.8in]{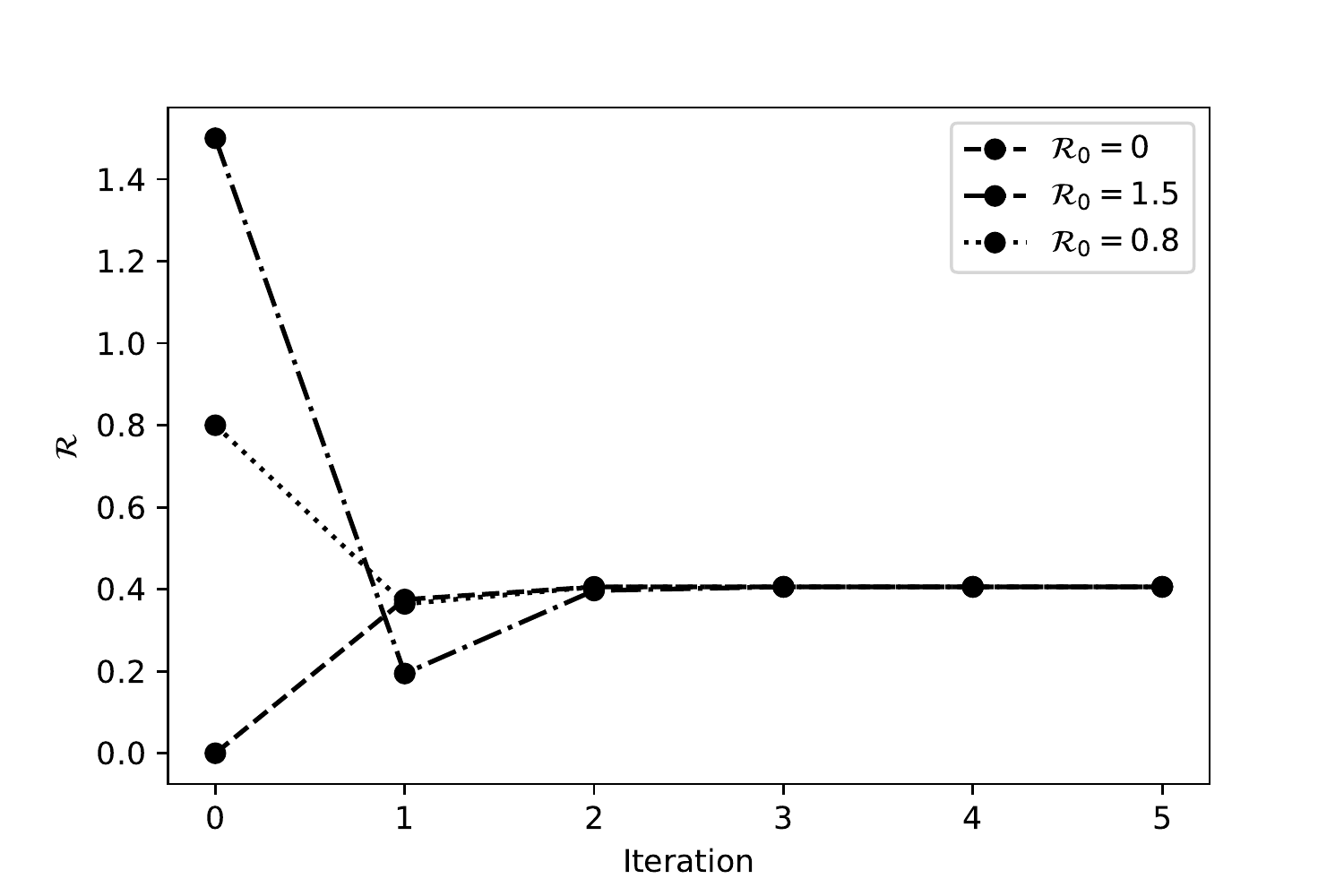}
\caption{Convergence of the iterative procedure to $\cR^*\approx 0.40589$ from different starting points}\label{fig:iteration}
\end{minipage}
\end{figure}

Thus for $\cR\leq 1$, we obtain, 
$$M(\cR) = \frac{(1-\cR^2)/4 + (4-\cR^2)/8}{1+ (2-\cR)/4 + (1-\cR)/2} =\frac{6-3\cR^2}{2(8-3\cR)}.$$
And for $\cR\in(1,2]$, we obtain,
$$M(\cR) = \frac{(4-\cR^2)/8}{1+ (2-\cR)/4} =  \frac{4-\cR^2}{2(6-\cR)}.$$
$M(\cR) = 0$ for $\cR>2$. The function $M(\cR)$ is plotted in Figure~\ref{fig:example}, where one can see that its unique fixed point is its maximizer. This is the point $\cR^*\approx 0.40589$. Thus the optimal hourly prices are $p^*_1 \approx (1+0.40589)/2 =  0.7029$ and $p^*_2 \approx (2+0.40589)/2 =  1.2029$. Figure~\ref{fig:iteration} shows the convergence of the iterative procedure to $\cR^*$ from different starting points. The convergence is quick (the plot shows 5 iterations). Observe that as expected, the output values are no larger than $\cR^*$ after the first iteration.

%In Section~\ref{apx:learning} in the Appendix, I outline a procedure to estimate near-optimal prices in case where the worker doesn't a priori know the distribution of hourly valuations for the different customer classes. 

\begin{remark}\label{rem:queue}
The above procedure of iteratively estimating the earning rate and using it as the opportunity cost to compute the optimal prices for each customer class doesn't generally converge to the optimal prices in the presence of queueing, or in any system where price discrimination across classes is necessary despite identically distributed preferences. This can be seen from the example presented in Section~\ref{sec:patient}. In that example, despite the customer classes having the identically distributed hourly valuations, I showed that price discrimination was necessary for optimality. But the above procedure can never converge to different prices for the two customer classes, since for any estimated earning rate, the optimal pricing problem solved by the procedure for the two classes in every iteration is identical. 
\end{remark}

%Thus, as the algorithm decreases the price until the earnings continue to increase, if $p''$ is the price at which the highest earning rate is observed and $p'$ is the price where the earnings fall for the first time, then any price lower than $p'$ cannot be the optimal price.

%One can moreover see that even when other classes are considered, the optimal earning rate cannot be lower than the optimal earning rate seen when this is the only available class. Hence, the optimal price for this class when other classes are considered can never be lower than the optimal price seen when this is the only available class. Thus there is no need for experimenting with lower prices. 

\section{Worker competition}\label{sec:competition}
A natural question that arises at this point is if any of these insights carry over in the presence of competition. 
Indeed, in reality, several workers offering the same services operate on any service platform. The traditional revenue management response to this concern would be that the customer arrival rates already factor in everything exogenous to the decisions of the worker in focus; in particular, these are the {\it residual} arrival rates after having accounted for the competition. However, such a response is valid only if one assumes that the competitive environment is stable, and that it does not change in response to the prices set by the worker. There is a priori no reason to assume so in our setting. In this section, I offer a rather optimistic resolution of this concern.

First, I start with a basic observation about a setting that I will refer to as {\it uniform workers setting}. Suppose that there are $N\geq 2$ workers that offer identical services, which are indistinguishable in terms of dimensions like quality, etc. Moreover, without loss of generality, suppose that their hourly service costs are identical and equal to $0$. Also, suppose that there is a single customer class with Poisson arrivals with rate $\lambda$, and an arbitrary job length distribution with a finite mean $1/\mu$. Each customer's maximum hourly willingness-to-pay is drawn independently from a strictly regular common distribution $F$. Moreover, suppose that each arriving customer, a)  chooses the cheapest available worker whose hourly price is lower than her hourly willingness to pay, b) leaves if there is no such worker or if all the workers are unavailable, and c) breaks ties arbitrarily if there are multiple workers that offer the same smallest price. Given a vector of prices $(p_1,\cdots, p_N)$ chosen by the workers, let $\cR_i(p_1,\cdots, p_N)$ be the long-run average revenue of worker $i$ for $i\in \{1,\cdots, N\}$. Recall the definition of Nash equilibrium.

\begin{definition}[Nash Equilibrium]
We say that the vector of prices $(p_1,\cdots, p_N)$ is a Nash equilibrium if for all $i\in \{1,\cdots, N\}$, 
\begin{align*}
\cR_i(p_1,\cdots, p_i,\cdots ,p_N)\geq \cR_i(p_1,\cdots, p'_i,\cdots ,p_N) \textrm{ for any } p'_i\in \mathbbm{R}.
\end{align*}
\end{definition}

I informally call such a Nash equilibrium vector of prices a {\it price equilibrium}. The main observation is that in the uniform workers setting, there exists no price equilibrium. The basic idea is the same as the ``Edgeworth Paradox'' in economics \cite{fisher1925edgeworth}. We record this observation in the following proposition.
%Essentially, as long as the prices of the workers are non-zero, all workers are incentivized to undercut each other by small amounts to obtain the largest share of the customer demand in exchange for negligible loss in per-customer revenue. This implies that any non-zero vector of prices cannot be an equilibrium. But even the zero vector is not a Nash equilibrium, since every worker is incentivized to raise the price to capitalize (i.e., earn a positive expected revenue) in the situations where all other workers are busy and they are the only worker available for the customer. We record this observation in the following proposition.

\begin{proposition}
In the uniform workers setting described above with $N\geq 2$, there exists no Nash equilibrium vector of prices.
\end{proposition}
\proof{Proof.}
Suppose that there exists a price equilibrium and without loss of generality, let workers $1$ and $2$ offer the two smallest prices at equilibrium, $p_1\leq p_2$. Let $p^*$ be the unique (due to strict regularity of $F$) monopolistic optimal price when there is a single worker. First, note that it cannot be that $p_1>p^*$, since if so then $1$ can lower his price to $p^*$ and make the monopolistic optimal revenue. Also, it cannot be that $p_1< p^*\leq p_2$, since again $1$ can raise her price closer to $p^*$ and make a revenue closer to the monopolistic optimal revenue. Thus it must be that $p_1\leq p_2\leq p^*$. It cannot be that $p_1<p_2\leq p^*$, since $1$ can raise her price while still being less than $p_2$ and obtain a higher revenue by inching closer to the monopolistic optimal price of $p^*$. Thus we have $p_1=p_2\leq p^*$; but again this cannot be an equilibrium unless $p_1=p_2 =0$, since either $1$ or $2$ can undercut the other and get a larger share of the market for negligible loss in revenue. But $p_1 = p_2=0$ cannot be an equilibrium either since either worker is incentivized to raise the price to capitalize (i.e., earn a positive expected revenue) in the situations where all the other workers are busy and they are the only worker available for the customer. Hence, there is no price equilibrium. \halmos

This suggests that amongst undifferentiated workers, there is no hope of obtaining a price equilibrium. However, it is rarely the case that the workers on a platform are undifferentiated -- they are typically ranked in order of quality as determined by the reputation systems of the platforms. In this case, I argue that the situation may not be as bleak and a price equilibrium may exist. Consider again the setting where there are $K$ classes of customers, with Poisson arrivals (rate $\lambda_k$) and arbitrary job length distributions (mean $1/\mu_k$).  Suppose that a class $k$ customer's hourly willingness to pay is drawn independently from the distribution $F_k$. There are $N$ workers, $\{1,\cdots,N\}$, ranked in descending order of perceived quality, i.e, $1>2>3>\cdots>N$. Each worker $i$ sets a price $p_{ik}$ for a customer of class $k$. 
I propose the following model for how a customer accounts for price and quality information in making purchase decisions on platforms.

\begin{definition}[A best-I-can-afford (BICA) buyer]
From the available workers, the customer chooses the highest ranked worker with an hourly price (for her class) less than or equal to her hourly willingness to pay. She leaves if there is no such worker.
\end{definition}

I argue that this is a natural model for quality sensitive customer behavior on platforms. For example, for high-value software development jobs on labor platforms like Upwork, clients have predetermined management approved budgets for their hourly expenditures and naturally, they want to get the best available freelancer that meets their budget constraint. Moreover, in cases where customers are not expected to persist on the platform for long, the absolute values of the worker reputation scores may not be very informative beyond the ranking information that they provide and thus, opting for the best ranked worker that one can afford may be perceived to be a safe strategy in the face of quality uncertainty.
%For example, there may not be a big difference between the projected experience of a 4.5 star restaurant and a 5 star restaurant on Yelp, but there is typically a big difference between the experience with a 4.5 star driver and a 5 star driver on Uber. For a discussion of such calibration issues, see \cite{garg2018designing}. 
Finally, the search and filtering abilities provided by platforms naturally support such behavior by making it easy to list products or services in decreasing order of quality scores. 
%On the other hand, this model may not be a good fit for platforms offering highly standardized services with little variation in perceived quality across workers. 
%making it easy for the customer to settle on the first worker on the list that fits the bill.

With this model, I argue that not only does a price equilibrium exist, it can be computed efficiently. The idea is simple. Since the highest rated worker 1 gets priority from the customers, she is unaffected by the pricing decisions of the remaining workers. That is, when she is available, she gets priority from all the customers who can afford to pay the price that she sets, irrespective of prices set by other available workers. Thus, she can independently solve the single worker price optimization problem that we solved in Section~\ref{sec:hetero}. Once her prices are fixed, she consumes a part of the incoming demand of customers who can afford to pay her prices -- not all since a fraction of these customers are rejected service from her because she is busy. The residual demand is thus composed of all the arriving customers who cannot afford to pay the first worker's prices and a fraction of the customers who are able to but are denied service due to unavailability. The second worker now has priority over these customers and is unaffected by the pricing decisions of the workers who have a lower rating than her. Thus, she again solves the single worker price optimization problem with the residual demand as earlier. Then the third worker sets the optimal prices with the residual demand that is rejected by the first and second workers and so on and so forth, until all the workers have set their prices. We record this observation in the following proposition.

\begin{proposition}
With workers ranked by their quality and BICA buyers, a price equilibrium exists.
\end{proposition}
The above argument also implies the following two points. First, with customers' hourly willingness to pay identically distributed across multiple classes, price discrimination due to systemic reasons is unnecessary for worker 1, and assuming that worker 1 doesn't price discriminate, such a price discrimination is unnecessary for worker 2, and so on. Second, if the workers asynchronously implement the algorithm discussed in Section~\ref{sec:hetero}, this would lead to each worker converging to the equilibrium price. This is trivial to see for the highest rated worker. But once the highest rated worker's prices have converged, the second highest rated worker will face a stable environment and her prices will converge, and so on and so forth. 

%Finally, we can also show that the prices are decreasing in the quality.

%It is also straightforward to see that in this case, the price equilibrium coincides with the optimal prices set by the platform in a centralized setting, where the goal is to maximize the total revenue across the workers. To see this, note that under the optimal pricing strategy of the platform, the top ranked worker must set her monopolistic optimal hourly prices for the different job classes, 
\section{Discussion and future directions }\label{sec:extapp}
{\bf Extensions.} Below are a couple of extensions that are subsumed by the model.
\begin{enumerate}
\item {\bf Hourly vs. per-job pricing.} Hourly pricing is equivalent to per-job pricing in terms of maximizing the long-run earnings assuming that a) the mean job duration is commonly known to the worker and the customer, and b) a customer's per-unit-time valuation is same as her per-job valuation divided by the mean duration. In this case, the optimal per job price is the same as the optimal hourly prices multiplied by the mean duration of the job. 
%Thus, even if the optimal hourly price is the same across classes, it would lead to different lump-sum payments because of differences in mean job durations. 

\item {\bf Commissions.} Typically, workers have to pay a fixed percentage of their revenue to the platform. In this case, if the worker charges $p$ per hour to the customer, her net revenue is $\beta p$ for some $\beta<1$. In this case, we can simply redefine $p_k$ to be the net revenue rate of the worker coming in from customer class $k$ and the probability that an incoming customer is willing to support that rate can be defined to be $\overline{F}'_k(p) = \overline{F}_k(p/\beta)$. Thus the analysis remains identical.
\end{enumerate}

{\bf Applicability.} Removed from the context of on-demand platforms, the present paper essentially studies the optimal pricing problem faced by a single server in stochastic system with no queueing. The rising on-demand nature of services in the gig economy is the main reason why this question is of pertinence now and is likely the reason why this specific sub-case hasn't received much attention earlier. 

As I mentioned in the introduction, I believe that the present formulation is the most applicable to on-demand labor markets for skilled or semi-skilled labor for jobs like graphic design, software development, application/website development etc. In these cases, the amount of time clients are willing to spend in hiring a worker is negligible compared to the duration of the job. Additionally, workers are not incentivized to multi-task, i.e., accept multiple jobs at the same time, for two reasons. One is that clients are time-sensitive and offering a short turnaround time is critical to winning a contract. Moreover, and perhaps more importantly, platforms like Upwork provide worker monitoring service (by taking worker's screen snapshots at random intervals) to the client to ensure that she is billed for exactly the hours that the workers spend working. This further diminishes the incentive to multi-task since there is no possibility of billing multiple clients for the same time spent. This also makes hourly billing mutually attractive to both the client and the worker as compared to fixed prices: the worker doesn't have to worry about working too much without compensation (high uncertainty in duration is typical for large projects), and platform monitoring ensures that the client doesn't overpay.

The present formulation can be applicable for pricing on rental marketplaces like Airbnb, Turo, etc., except for a couple of caveats. In these cases, there is seasonality in demand, e.g., weekends demand characteristics may be different from weekdays, summer demand may have different characteristics than other seasons, etc. Although these can be modeled as different customer classes, due to the cyclic nature of their arrivals, Poisson arrivals may not be an appropriate assumption. An additional concern is that hosts may want to price discriminate based on the time from the date of renting, similar to traditional revenue management settings, since high paying customers typically arrive later. Having said so, either because of the desire for simplicity of the price menu, or because these above two effects are believed to be negligible, the optimal pricing formulation presented in the paper can potentially be a reasonable approximation in these settings. 

%Assuming there is little spillover, the worker can potentially solve the optimal pricing problem for different periods independently using the formulation from the present work. 

%The flexibility and autonomy of freelance work on service platforms is accompanied by the significant risk resulting from the lack of stability and benefits associated with full time employment. Effectively mitigating these risks is critical to the long-run success of the gig economy. Understanding and addressing the different operational challenges that workers face can not only directly improve short-term outcomes for the workers, but it can also help the platforms predict overall system behavior and design effective controls that improve efficiency in the long-term. 

{\bf Future directions.} There are several interesting avenues for further investigation. It would be interesting to quantify the impact of decentralized competitive pricing on platform revenues when compared to centralized pricing. 
Another issue is that of learning customer preferences. In the present work, I assumed that the worker knows the distribution of the customers' per-unit-time willingness to pay. It would be interesting to explore how these distributions can be learned while strategically experimenting with prices. 
%This aligns with the theme of pricing under demand learning that has been popular in the operations management literature in the recent years. 
The novel aspect here is the need to account for externalities imposed by accepting jobs on the intertwined goals of revenue maximization and demand learning. I am optimistic that the structure of the optimal pricing solution and especially the iterative procedure that I presented in this paper will inform the design of good ``learning while earning'' pricing strategies in this setting. 

\bibliographystyle{apalike}
\bibliography{ondemand}

\begin{thebibliography}{}

\bibitem[Armstrong, 2006]{armstrong2006competition}
Armstrong, M. (2006).
\newblock Competition in two-sided markets.
\newblock {\em The RAND Journal of Economics}, 37(3):668--691.

\bibitem[Bakos and Katsamakas, 2008]{bakos2008design}
Bakos, Y. and Katsamakas, E. (2008).
\newblock Design and ownership of two-sided networks: Implications for internet
  platforms.
\newblock {\em Journal of Management Information Systems}, 25(2):171--202.

\bibitem[Banerjee et~al., 2015]{banerjee2015}
Banerjee, S., Johari, R., and Riquelme, C. (2015).
\newblock Pricing in ride-sharing platforms: A queueing-theoretic approach.
\newblock In {\em Proceedings of the Sixteenth ACM Conference on Economics and
  Computation}, EC '15.

\bibitem[Bimpikis et~al., 2019]{bimpikis2019spatial}
Bimpikis, K., Candogan, O., and Saban, D. (2019).
\newblock Spatial pricing in ride-sharing networks.
\newblock {\em Operations Research}.

\bibitem[Birge et~al., 2018]{birge2018optimal}
Birge, J., Candogan, O., Chen, H., and Saban, D. (2018).
\newblock Optimal commissions and subscriptions in networked markets.
\newblock In {\em Proceedings of the 2018 ACM Conference on Economics and
  Computation}, pages 613--614. ACM.

\bibitem[Cachon et~al., 2017]{cachon2017role}
Cachon, G.~P., Daniels, K.~M., and Lobel, R. (2017).
\newblock The role of surge pricing on a service platform with self-scheduling
  capacity.
\newblock {\em Manufacturing \& Service Operations Management}, 19(3):368--384.

\bibitem[Caillaud and Jullien, 2003]{caillaud2003chicken}
Caillaud, B. and Jullien, B. (2003).
\newblock Chicken \& egg: Competition among intermediation service providers.
\newblock {\em RAND journal of Economics}, pages 309--328.

\bibitem[Caro and Simchi-Levi, 2012]{caro2012optimal}
Caro, F. and Simchi-Levi, D. (2012).
\newblock Optimal static pricing for a tree network.
\newblock {\em Annals of Operations Research}, 196(1):137--152.

\bibitem[Castillo et~al., 2017]{castillo2017surge}
Castillo, J.~C., Knoepfle, D., and Weyl, G. (2017).
\newblock Surge pricing solves the wild goose chase.
\newblock In {\em Proceedings of the 2017 ACM Conference on Economics and
  Computation}, pages 241--242. ACM.

\bibitem[De~Stefano, 2015]{de2015rise}
De~Stefano, V. (2015).
\newblock The rise of the just-in-time workforce: On-demand work, crowdwork,
  and labor protection in the gig-economy.
\newblock {\em Comp. Lab. L. \& Pol'y J.}, 37:471.

\bibitem[Economides and Katsamakas, 2006]{economides2006two}
Economides, N. and Katsamakas, E. (2006).
\newblock Two-sided competition of proprietary vs. open source technology
  platforms and the implications for the software industry.
\newblock {\em Management Science}, 52(7):1057--1071.

\bibitem[Fisher, 1925]{fisher1925edgeworth}
Fisher, I. (1925).
\newblock Edgeworth's papers relating to political economy.
\newblock {\em The Quarterly Journal of Economics}, 40(1):167--171.

\bibitem[Gallager, 2013]{gallager2013stochastic}
Gallager, R.~G. (2013).
\newblock {\em Stochastic processes: theory for applications}.
\newblock Cambridge University Press.

\bibitem[Hassin and Haviv, 2003]{hassin2003queue}
Hassin, R. and Haviv, M. (2003).
\newblock {\em To queue or not to queue: Equilibrium behavior in queueing
  systems}, volume~59.
\newblock Springer Science \& Business Media.

\bibitem[Lin et~al., 2011]{lin2011innovation}
Lin, M., Li, S., and Whinston, A.~B. (2011).
\newblock Innovation and price competition in a two-sided market.
\newblock {\em Journal of Management Information Systems}, 28(2):171--202.

\bibitem[Myerson, 1981]{myerson1981optimal}
Myerson, R.~B. (1981).
\newblock Optimal auction design.
\newblock {\em Mathematics of Operations Research}, 6(1):58--73.

\bibitem[Olive, 2008]{olive2008applied}
Olive, D.~J. (2008).
\newblock Applied robust statistics.

\bibitem[Rochet and Tirole, 2003]{rochet2003platform}
Rochet, J.-C. and Tirole, J. (2003).
\newblock Platform competition in two-sided markets.
\newblock {\em Journal of the european economic association}, 1(4):990--1029.

\bibitem[Rochet and Tirole, 2006]{rochet2006two}
Rochet, J.-C. and Tirole, J. (2006).
\newblock Two-sided markets: a progress report.
\newblock {\em The RAND journal of economics}, 37(3):645--667.

\bibitem[Taylor, 2018]{taylor2018demand}
Taylor, T.~A. (2018).
\newblock On-demand service platforms.
\newblock {\em Manufacturing \& Service Operations Management}, 20(4):704--720.

\bibitem[Weyl, 2010]{weyl2010price}
Weyl, E.~G. (2010).
\newblock A price theory of multi-sided platforms.
\newblock {\em American Economic Review}, 100(4):1642--72.

\bibitem[Ziya et~al., 2006]{ziya2006optimal}
Ziya, S., Ayhan, H., and Foley, R.~D. (2006).
\newblock Optimal prices for finite capacity queueing systems.
\newblock {\em Operations Research Letters}, 34(2):214--218.

\bibitem[Ziya et~al., 2008]{ziya2008note}
Ziya, S., Ayhan, H., and Foley, R.~D. (2008).
\newblock A note on optimal pricing for finite capacity queueing systems with
  multiple customer classes.
\newblock {\em Naval Research Logistics (NRL)}, 55(5):412--418.

\end{thebibliography}

\begin{APPENDIX}{}
\section{Derivation of \eqref{hanger}}
Let $P_0$ be the expected earnings between two successive times when the system is empty and the worker is idle (expected earnings in a renewal cycle). Let $P_1$ ($P_2$) be the expected earning till the system is empty starting from the time when a job of class A (B) is accepted and there is no customer in queue. Let $T_0$ be the expected time till the system is empty again after taking up a job, starting with an empty system (expected duration of a renewal cycle). Let $T_1$ ($T_2$) be the expected time till the system is empty starting from the time when a job of class A (B) is accepted and there is no customer in queue. Then from the renewal reward theorem, the long-run average earning converges almost surely to $P_0/T_0$. Denote $\lambda'_i= \lambda_i(1-p_i)$ for $i=1,\,2$. Then we have the following set of first-step equations satisfied by the various quantities.
\begin{align}
P_0 &=\frac{\lambda'_1P_1}{\lambda'_1+\lambda'_2} + \frac{\lambda'_2P_2}{\lambda'_1+\lambda'_2}\\
P_1 &= \frac{p_1}{\mu_1}+\frac{\lambda'_1P_1}{\lambda'_1+\lambda'_2+\mu_1} + \frac{\lambda'_2P_2}{\lambda'_1+\lambda'_2+\mu_1}\\
P_2 &= \frac{p_2}{\mu_2}+\frac{\lambda'_1P_1}{\lambda'_1+\lambda'_2+\mu_2} + \frac{\lambda'_2P_2}{\lambda'_1+\lambda'_2+\mu_2}\\
T_0 &=\frac{1}{\lambda'_1+\lambda'_2}+\frac{\lambda'_1T_1}{\lambda'_1+\lambda'_2} + \frac{\lambda'_2T_2}{\lambda'_1+\lambda'_2}\\
T_1 &= \frac{1}{\mu_1}+\frac{\lambda'_1T_1}{\lambda'_1+\lambda'_2+\mu_1} + \frac{\lambda'_2T_2}{\lambda'_1+\lambda'_2+\mu_1}\\
T_2 &= \frac{1}{\mu_2}+\frac{\lambda'_1T_1}{\lambda'_1+\lambda'_2+\mu_2} + \frac{\lambda'_2T_2}{\lambda'_1+\lambda'_2+\mu_2}.
\end{align}
Solving, we obtain \eqref{hanger} as the expression for $P_0/T_0$.
\end{APPENDIX}

%%%%%%%%%%%%%%%%%
\end{document}